\documentclass{edm_article}

\usepackage{url}
\usepackage{quoting}
\usepackage{booktabs}
\usepackage{multirow}
\quotingsetup{vskip=0pt}
\begin{document}

\title{Towards Modeling Learner Performance with Large Language Models}

% Submissions for EDM are double-blind: please do not include any author names or affiliations in the submission. 
% Anonymous authors:
% \numberofauthors{1}
% \author{
% Anonymous\\
%        \affaddr{Anonymous Institution}\\
%        \email{anonymous@anonymous.edu}
% }
%An example of how to include
% multiple authors is below for after the paper has been accepted.


\numberofauthors{6} %  in this sample file, there are a *total*
% % of EIGHT authors. SIX appear on the 'first-page' (for formatting
% % reasons) and the remaining two appear in the \additionalauthors section.
% %
\author{
% 1st. author
\alignauthor
Seyed Parsa Neshaei\thanks{Equal contribution}\\
       \affaddr{EPFL}\\
       \email{seyed.neshaei@epfl.ch}
% 2nd. author
\alignauthor
Richard Lee Davis\footnotemark[1]\\
       \affaddr{EPFL}\\
       \email{richard.davis@epfl.ch}
% 3rd. author
\alignauthor Adam Hazimeh\\
       \affaddr{EPFL}\\
       \email{adam.hazimeh@epfl.ch}
\and  % use '\and' if you need 'another row' of author names
% 4th. author
\alignauthor Bojan Lazarevski\\
       \affaddr{EPFL}\\
       \email{bojan.lazarevski@epfl.ch}
% 5th. author
\alignauthor Pierre Dillenbourg\\
       \affaddr{EPFL}\\
       \email{pierre.dillenbourg@epfl.ch}
% 6th. author
\alignauthor Tanja Käser\\
       \affaddr{EPFL}\\
       \email{tanja.kaeser@epfl.ch}
}
% % There's nothing stopping you putting the seventh, eighth, etc.
% % author on the opening page (as the 'third row') but we ask,
% % for aesthetic reasons that you place these 'additional authors'
% % in the \additional authors block, viz.
% \additionalauthors{Additional authors: John Smith (The Th{\o}rv{\"a}ld Group,
% email: {\texttt{jsmith@affiliation.org}}) and Julius P.~Kumquat
% (The Kumquat Consortium, email: {\texttt{jpkumquat@consortium.net}}).}
% \date{30 July 1999}
% Just remember to make sure that the TOTAL number of authors
% is the number that will appear on the first page PLUS the
% number that will appear in the \additionalauthors section.

\maketitle

\begin{abstract}
Recent work exploring the capabilities of pre-trained large language models (LLMs) has demonstrated their ability to act as general pattern machines by completing complex token sequences representing a wide array of tasks, including time-series prediction and robot control. This paper investigates whether the pattern recognition and sequence modeling capabilities of LLMs can be extended to the domain of knowledge tracing, a critical component in the development of intelligent tutoring systems (ITSs) that tailor educational experiences by predicting learner performance over time. In an empirical evaluation across multiple real-world datasets, we compare two approaches to using LLMs for this task, zero-shot prompting and model fine-tuning, with existing, non-LLM approaches to knowledge tracing. While LLM-based approaches do not achieve state-of-the-art performance, fine-tuned LLMs surpass the performance of naive baseline models and perform on par with standard Bayesian Knowledge Tracing approaches across multiple metrics. These findings suggest that the pattern recognition capabilities of LLMs can be used to model complex learning trajectories, opening a novel avenue for applying LLMs to educational contexts. The paper concludes with a discussion of the implications of these findings for future research, suggesting that further refinements and a deeper understanding of LLMs’ predictive mechanisms could lead to enhanced performance in knowledge tracing tasks\footnote{Our code and links to data are available on \url{https://github.com/spneshaei/Towards-Modeling-Learner-Performance-with-LLMs}}.

\end{abstract}

\keywords{large language models, knowledge tracing, student modeling} % Replace with your own 3-5 keywords

\section{Introduction}
\label{sec:introduction}
Pre-trained large language models (LLMs), such as BERT \cite{devlin2018bert} and GPT-3 \cite{brown2020language}, are transformer-based neural networks containing large numbers of parameters (e.g., from 110 million in BERT to 175 billion in GPT-3) trained on massive amounts of natural language data.
LLMs have achieved state-of-the-art results on multiple tasks, including natural language modeling, question answering, sentiment analysis, and common sense reasoning \cite{brown2020language}.
These models demonstrate an impressive ability for zero-shot and few-shot learning, which is the ability to generalize to novel tasks when provided with a handful of examples or task instructions in natural language.
For example, LLMs have matched or exceeded the performance of bespoke models on several benchmarks, such as computer programming \cite{chenEvaluatingLargeLanguage2021}, medicine \cite{thirunavukarasu2023large}, and mathematics \cite{lewkowyczSolvingQuantitativeReasoning2022}, demonstrating their ability to leverage learned language patterns across a variety of domains.

In the domain of education, one of the emerging uses of LLMs is integration into intelligent tutoring systems (ITSs), which are digital learning environments designed to provide intelligent assistance to students in the form of hints, explanations, and adaptive instructional content \cite{graesser2012intelligent, anderson1985intelligent}.
% Add one more sentence explaining what ITSs provide, to help make the transition to LLMs.
Recent work suggests that LLMs are capable of providing personalized learning experiences to students in multiple domains \cite{su2023reviewriter} and in various curricula and topics, including programming \cite{cao2023scaffolding}, biology \cite{sanpablo2023development}, math \cite{zong2023solving}, or chemistry \cite{white2023assessment}.
% SOMEDAY: Explain the roles that LLMs play/support in ITSs.

So far, the actual work of modeling students' knowledge and performance has not been handled by LLMs directly.
Instead, this task is specialized accomplished by knowledge tracing models that track how students learn over time through their interactions with the ITS.
Knowledge tracing models are typically applied to the question-answering part of an ITS and model the evolution of student knowledge based on their responses to previous questions.
This enables the models to make predictions about how a student would answer subsequent questions and to adaptively select questions to present to the learner in their exercise at any given stage, with the goal of providing a relevant curriculum of materials to maximize learning gains \cite{abdelrahman2023knowledge, corbett1994knowledge, schodde2017adaptive}.

Knowledge tracing models in the EDM literature can be categorized broadly into distinct families, each with unique approaches to modeling student learning. Approaches based on Markov models like Bayesian Knowledge Tracing (BKT) \cite{anderson1985intelligent} form the first family, while approaches based on logistic regression such as Learning Factors Analysis (LFA) \cite{cenLearningFactorsAnalysis2006a} form a second family. Both families model student learning with a limited set of parameters that correspond to mastery levels, student ability, task characteristics, learning rates, and other factors depending on the model.

A third family utilizes deep neural networks to model skill acquisition and learner performance. The foundational approach in this family, known as Deep Knowledge Tracing (DKT) \cite{piech2015deep}, uses recurrent neural networks to process student interaction data and jointly model skill acquisition. Subsequent approaches in this family have added dynamic memory networks \cite{zhangDynamicKeyValueMemory2017}, exercise content \cite{suExerciseEnhancedSequentialModeling2018,liuEKTExerciseAwareKnowledge2021}, and self-attention \cite{choiAppropriateQueryKey2020}, resulting in improved interpretability and moderate gains in performance over DKT in some cases \cite{abdelrahman2023knowledge}.

The goal of the current paper is to explore whether LLMs might serve as the basis for a new family of methods for student performance prediction.
Our inspiration is based on recent work demonstrating that LLMs can act as general pattern machines, learning to model and predict abstract sequences of numerical data encoded as strings of natural language tokens \cite{mirchandaniLargeLanguageModels2023}.
%SOMEDAY: Explain what the benefits of this approach, or at least try to explain why this is interesting

Our primary research question is:
\begin{quoting}
    \textbf{RQ1:} Can the general pattern matching abilities of LLMs be applied to the domain of student performance prediction?
\end{quoting}
To answer this question, we evaluate two distinct approaches to applying LLMs in this domain, zero-shot prediction and model fine-tuning. For each approach, we ask the question:
\begin{quoting}
    \textbf{RQ2:} How does the performance of the LLM on three real-world datasets compare to (a) naive baselines, (b) Markov approaches to knowledge tracing, (c) logistic regression approaches, and (d) deep learning approaches to knowledge tracing?
\end{quoting}
Finally, in the context of fine-tuning, we investigate the question:
\begin{quoting}
    \textbf{RQ3:} Is there a relationship between the scale of a fine-tuned model (in terms of parameter count) and its ability to predict student performance?
\end{quoting}

In the current work, we selected three datasets from the literature which were previously used in knowledge tracing research.
We evaluated our two LLM-based approaches on the data, as well as comparing them with naive baselines and a range of standard knowledge tracing approaches across a set of metrics.
We found while the zero-shot approaches were not successful in capturing student knowledge over time, fine-tuned LLMs beat the baselines and exhibited similar or higher performance than standard Bayesian models for knowledge tracing.
However, the fine-tuned LLMs did not manage to surpass the performance of other approaches for knowledge tracing, such as DKT.
Our results shed light on the applicability of LLMs as general pattern machines to the task of knowledge tracing, and enhance EDM research by contributing to the research line of the effectiveness of LLMs in pedagogical scenarios.

\section{Related Work}

\subsection{Approaches to Student Performance Modeling}

There are many approaches to implementing knowledge tracing models\footnote{For a comprehensive survey, see \cite{abdelrahman2023knowledge}.} that can be categorized into a smaller number of families.
The first family of approaches models student knowledge using Markov models, as typified by Bayesian Knowledge Tracing (BKT) \cite{anderson1985intelligent}.
BKT models student learning as a binary process, where knowledge components are either mastered or not, and updates students' mastery levels based on their interactions with tasks, utilizing four parameters to account for prior knowledge, learning transitions, and the probabilities of guessing or slipping.
Variations of BKT have been developed to incorporate additional dimensions such as student assistance received \cite{linInterventionBKTIncorporatingInstructional2016}, task difficulty \cite{pardosKTIDEMIntroducingItem2011}, and individualized adjustments to learning and error probabilities \cite{yudelson2013individualized}, enhancing the model's ability to personalize learning assessments and make predictions.

A second family of approaches uses logistic regression to model learning and predict performance.
Logistic regression predictions take the form $$p(a_{s,t+1}=1|q_{s,t+1},\textbf{x}_{s,1:t})=\sigma(\textbf{w}^T\Phi(q_{s,t+1},\textbf{x}_{s,1:t}))$$ where $a_{s,t+1}$ is the response prediction of learner $s$ at timestep $t+1$, $q_{s,t+1}$ is the question of learner $s$ at timestep $t+1$, $\textbf{x}_{s,1:t}$ is all of the data for learner $s$ up to timestep $t$, $\sigma$ is the logistic function, $\textbf{w} \in \mathbb{R}^d$ is a trainable weight vector, and $\Phi(q_{s,t+1},\textbf{x}_{s,1:t})$ is is a vector of $d$ features representing the learner, question $q_{t+1}$, and history $\textbf{x}_{1:t}$.
Different feature vectors $\Phi$ are associated with different models.
For example, Performance Factors Analysis (PFA) \cite{pavlikjrPerformanceFactorsAnalysis2009} creates a feature vector that includes the difficulty of each knowledge component $k$ associated with question $q_{t+1}$ and the number of correct and incorrect answers on each knowledge component $k$ by learner $s$ prior to timestep $t+1$.
Other approaches in this family include Learning Factors Analysis (LFA) \cite{cenLearningFactorsAnalysis2006a}, DAS3H (which includes continuous time window features) \cite{choffinDAS3HModelingStudent2019}, and Best-LR (which is similar to PFA but includes the total prior number of correct and wrong responses in the features) \cite{gervet2020deep}.

A third family of knowledge tracing approaches utilizes deep neural networks to model student performance.
This approach was pioneered by Piech et al. with the Deep Knowledge Tracing (DKT) model \cite{piech2015deep}, which used a recurrent neural network architecture to process student activity and jointly model skill acquisition at each timestep.
Recent advancements in knowledge tracing include the introduction of dynamic memory models like DKVMN \cite{zhangDynamicKeyValueMemory2017} and SKVMN \cite{abdelrahmanKnowledgeTracingSequential2019}, which incorporate next skill input and LSTM integration, respectively, the utilization of exercise textual content in EERNN \cite{suExerciseEnhancedSequentialModeling2018} and EKT \cite{liuEKTExerciseAwareKnowledge2021} models, and the exploration of self-attentive transformer models for knowledge tracing, as demonstrated by the SAKT model \cite{pandey2019self}.

To the best of our knowledge, our work is the first to investigate whether LLMs might serve as the basis for a new family of knowledge tracing models.
In this proposed family, student interaction data would be represented by sequences of natural language tokens, and a complete dataset would form a kind of specialized sub-language.
Modeling and prediction tasks would rely on an LLM’s ability to recognize or learn the underlying linguistic structure in these sub-languages and extrapolate from seen to unseen data.
This approach to student performance prediction is inspired by recent work showing that, under the right conditions, LLMs are capable of identifying patterns in sequences of tokens representing numerical data and predicting unseen behavior.

\subsection{LLMs as General Pattern Machines}
While LLMs excel at a variety of tasks involving natural language, there is a growing interest in their abilities to act as general pattern machines \cite{mirchandaniLargeLanguageModels2023} by modeling and making predictions about more general, non-linguistic patterns.
A general example of this approach is LLMTime \cite{gruverLargeLanguageModels2023}, which uses pre-trained LLMs for zero-shot, continuous time prediction of numerical data.
By representing the time series data as a string of numerical digits that can be properly tokenized, this method is able to treat time-series forecasting as next-token prediction in text, achieving high predictive accuracy across multiple datasets.
Other successful approaches to using LLMs for general time-series modeling and prediction have utilized alternative prompting schemes such as time-series decomposition \cite{caoTEMPOPromptbasedGenerative2023} or fine-tuning on sequence data \cite{zhouOneFitsAll2023,changLLM4TSAligningPreTrained2024}.

LLMs have also been used for time-series prediction in specific domains, including healthcare \cite{chenMEDITRON70BScalingMedical2023,liuLargeLanguageModels2023}, finance \cite{yu2023temporal,lopez2023can}, and transportation \cite{xueLeveragingLanguageFoundation2022}.
However, to the best of our knowledge, there has been no work investigating the abilities of LLMs to model and predict time-series data about student performance in educational contexts.
The goal of the current work is to explore the suitability of two common approaches to using LLMs for time-series prediction---zero-shot prediction and fine-tuning---on the task of student performance prediction.
We provide more details on these approaches in the following sections.

\section{Evaluated Models}
\label{sec:evaluated-models}

To compare how LLMs can reliably predict the answers to the questions and thus help in knowledge tracing, we compared their performance against well-established traditional approaches to knowledge tracing.

\subsection{Problem Formulation}

To model students' knowledge over time, we followed the approach proposed by Gervet et al. \cite{gervet2020deep}. We define $S$ as the sequence of answers provided by a student, and $S_{:i}$ as the sub-sequence of answers provided up to and before point $i$ in sequence $S$. We also define $K_i$ as the ID indicating the skill (i.e., knowledge component) of the current next question. At each point $i$, we aim to predict the correctness of the student response in the next step, using data from $S_{:i}$. For simplicity, we consider \textit{correct} answer as 1, and \textit{wrong} answer as 0.

\subsection{Baselines and Previous Models}

For our naive baselines, we used three approaches:

\begin{itemize}
    \item \textbf{Mean:} Following \cite{sarsa2021empirical}, we found the mean of all the responses in any given dataset and constantly predicted the same value as a simple baseline. The mean is computed over all of the training data rather than specific users or knowledge components, for simplicity \cite{sarsa2021empirical}. We also computed the (constant) probability of each prediction as the count of responses equal to the mean response, divided by the total number of responses in the whole training dataset.

    \item \textbf{Next as Previous (NaP):} Extending the approach used by \cite{sarsa2021empirical}, at each step $i$, we considered the mean of all previous responses in $S_{:i}$ as the next prediction. We also computed the probability of each prediction as the count of responses equal to the mean response in $S_{:i}$, divided by the total number of responses in $S_{:i}$.

    \item \textbf{Next as Previous, among those having the same skill (NaP Skills):} We also extended the NaP baseline by only considering the mean of the previous responses in $S_{:i}$ with the same skill ID as of the current next question ($K_i$), as the next prediction. We also computed the probability of each prediction as the count of responses in $S_{:i}$ equal to the mean response and with skill ID $K_i$, divided by the total number of responses in $S_{:i}$ with skill ID $K_i$.
\end{itemize}

Additionally, we experimented with four other previously-used approaches in knowledge tracing, proven to be useful in the literature, and described briefly in the following:

\begin{itemize}
    \item \textbf{Bayesian Knowledge Tracing (BKT):} In its simplest form, a standard BKT HMM (Hidden Markov Model) aims to model the knowledge of the student as a latent variable, which is determined by four parameters per each skill: A) prior learning, B) probability of moving from the \textit{not-knowledgable} state to \textit{knowledgable} state for a skill, C) probability of applying a skill correctly by accident, and D) probability of applying a skill incorrectly by accident \cite{sarsa2021empirical, corbett1994knowledge, yudelson2013individualized}.

    \item \textbf{Deep Knowledge Tracing (DKT):} This model works with an RNN-based long short-term memory (LSTM) deep architecture. The \textit{memory} of the LSTM enables this architecture to model students attempting to solve questions with different skills over time and to predict whether the next question would be answered correctly or not. This architecture helps to reduce information loss due to the integrated gate structure \cite{sarsa2021empirical, piech2015deep, hochreiter1997long, hochreiter1996lstm}.

    \item \textbf{Best-Logistic Regression (Best-LR):} Introduced by \cite{gervet2020deep}, this approach uses a logistic regression model trained on the total number of previous success and error counts in $S_{:i}$ at each step $i$. They found that their approach beats the DKT model in several knowledge tracing datasets.

    \item \textbf{Self-Attentive Knowledge Tracing (SAKT):} This model, first introduced by \cite{pandey2019self}, uses a Transformer-based architecture \cite{vaswani2017attention} which operates using an embedding of information at point $i$ and predicts a probability for correctness at each step using a one-neuron output layer at the end.
\end{itemize}

\subsection{Fine-tuning LLMs}

We constructed a set of \textit{prompt-completion} pairs to fine-tune LLMs for the task of knowledge tracing. Each \textit{prompt} includes the details needed for prediction at step $i$, and the \textit{completion} includes the answer to be predicted for the current question at each step.

Our approach to constructing input prompts was directly inspired by logistic regression approaches to knowledge tracing \cite{gervet2020deep} which convert student interaction data into a set of features $\Phi(q_{s,t+1},\textbf{x}_{s,1:t})$.
Where our approach differs is that we represent this set of features $\Phi$ using natural language, and pass the set of natural language features to an LLM which then generates a prediction.

Thus, for our \textit{prompts}, we used two sets of features as inputs to our models, also among the feature sets used by \cite{gervet2020deep}, for each point $i$:

\begin{itemize}
    \item \textbf{1) Minimal Prompt:} This prompt includes the ID of the current question ($= A$), the total number of correct answers for all prior questions in $S_{:i}$ ($= B$), and the total number of wrong answers for all prior questions in $S_{:i}$ ($= C$). We thus formed our \textit{prompt} at each step $i$ accordingly as:

    "\textit{Total correct until now: $B$} \\
    \textit{Total wrong until now: $C$} \\
    \textit{Current question ID: $A$} \\
    \textit{Student response:} "

    and \textit{completion} as either "CORRECT" or "WRONG".
    
    \item \textbf{2) Extended Prompt:} This prompt includes all of the features in the minimal prompt, as well as the following features: $K_i$, the total number of correct answers for all prior questions in $S_{:i}$ with skill ID equal to $K_i$ ($= D$), and the total number of wrong answers for all prior questions in $S_{:i}$ with skill ID equal to $K_i$ ($= E$). We thus formed our \textit{prompt} at each step $i$ accordingly as:

    "\textit{Current skill ID: $K_i$} \\
    \textit{Total correct for prior questions with skill ID $K_i$: $D$} \\
    \textit{Total wrong for prior questions with skill ID $K_i$: $E$} \\
    \textit{Total correct until now: $B$} \\
    \textit{Total wrong until now: $C$} \\
    \textit{Current question ID: $A$} \\
    \textit{Student response:} "

    and \textit{completion} as either "CORRECT" or "WRONG".
\end{itemize}

We split the digits of all of the numbers used in our prompts by adding a space between each two consecutive digits (for example, "342" was changed to "3 4 2"), as this has been shown to improve the performance of LLMs when making predictions based on numerical data \cite{gruverLargeLanguageModels2023}.

We experimented using three different LLMs with different model sizes to observe the effect of larger models on their performance in the task of knowledge tracing:

\subsubsection{Fine-tuning BERT}
The Bidirectional Encoder Representations from Transformers (BERT) architecture \cite{devlin2018bert} is a Transformer-based model with around 109 million parameters trained using the masked language modeling approach, and commonly used for text classification \cite{gonzalez2020comparing, sun2019fine}, especially in educational applications to provide automated feedback to teachers or students \cite{jensen2021deep, weber2023structured, wulff2023utilizing}. We fine-tuned the \texttt{bert-base-cased} model provided by HuggingFace on our data for 3 epochs with a batch size of 32 and Adam optimizer having an initial learning rate of \texttt{3e-5}. We provided the \textit{prompts} as inputs to the model, and the binarized version of the \textit{completions} (1 for "CORRECT" and 0 for "WRONG") as labels per each input sentence.

\subsubsection{Fine-tuning GPT-2}
The Generative Pre-trained Transformer 2 (GPT-2) model \cite{radford2019language}, consisting of around 1.5 billion parameters, is capable of generating text completions, given partial text as input, and has been also used in studies on educational technologies \cite{su2023reviewriter, shi2022effidit, vinitra-thiemo-bias, tsai2021short}. We used GPT-2 as provided by HuggingFace and fine-tuned it on our data for 2 epochs with a batch size of 4. As input data, we provided concatenated \textit{prompts} and \textit{completions}, and separated \textit{prompt}-\textit{completion} pairs belonging to each training data point using an \texttt{<|endoftext|>} special token.

\subsubsection{Fine-tuning GPT-3}
Introduced by Brown et al. \cite{brown2020language}, this model, consisting of 175 billion parameters, significantly improves upon GPT-2 by exhibiting pattern matching and few-show learning capabilities \cite{brown2020language}, opening the way for utilizing \textit{prompt engineering} more effectively as a method to utilize the vast amount of knowledge that LLMs have acquired from their training data \cite{muktadir2023brief}. GPT-3 has also been used in numerous educational applications and intelligent tutoring systems \cite{cao2023scaffolding, abdelghani2023gpt, kikalishvili2023unlocking, tack2022ai}. We fine-tuned \texttt{babbage-002}, a GPT-3 base model developed by OpenAI, on our training data using the interface provided by OpenAI with default settings. We used the \texttt{logprobs} functionality in OpenAI's API to calculate the probabilities for output tokens. We computed the probabilities of each of the words CORRECT and WRONG individually by summing probabilities of their starting tokens (e.g., "C" and "COR" for the word CORRECT) extracted from the \texttt{logprobs} data. We then normalized the values by dividing each of the two probabilities over their sum.

\subsection{Zero-shot Prediction with GPT-3.5}
GPT-3.5 is a relatively recent model developed by OpenAI, trained using the Reinforcement Learning with Human Feedback (RLHF) approach, and incorporated in ChatGPT\footnote{\url{https://chat.openai.com}} to support users by providing answers in the form of an intelligent conversational agent \cite{wu2023brief}. Researchers have also used GPT-3.5 in \textit{zero-shot} settings, in which they do not fine-tune the model, but rather ask the model to give an answer based on a single self-describing input prompt with no other contextual input \cite{wei2023zero, wang2023zero}.

We used the same minimal prompt\footnote{Due to resource limitations, we did not conduct our zero-shot experiment on extended prompts, and leave it for future work.}, as used in our fine-tuning process, for the \textit{user message} of the prompt at each step $i$.
% , but removed the current question ID, as without fine-tuning on the training set, the question IDs had no meaning for the model. Thus, we only input the total number of \textit{correct} and \textit{wrong} answers in $S_{:i}$ for the \textit{user message} of the prompt at each step $i$.
Furthermore, to define the task for GPT-3.5 and instruct it to reply correctly given no training data, we included the following as the initial \textit{system message} in each prompt:

\textit{"You are an instructor and want to predict whether a student will get a question CORRECT or WRONG. The only information you have is the student's previous answers to a series of related questions. You know how many questions they got CORRECT and how many they got WRONG. Based on this information, you should make a prediction by outputting a single word: CORRECT if you think the student will answer the next question correctly, and WRONG if you think the student will answer the next question wrong. Output no other word at all, this is very important. Try to estimate the knowledge of the student before making your prediction."}

\begin{figure*}
    \centering
    \includegraphics[width=0.87\textwidth]{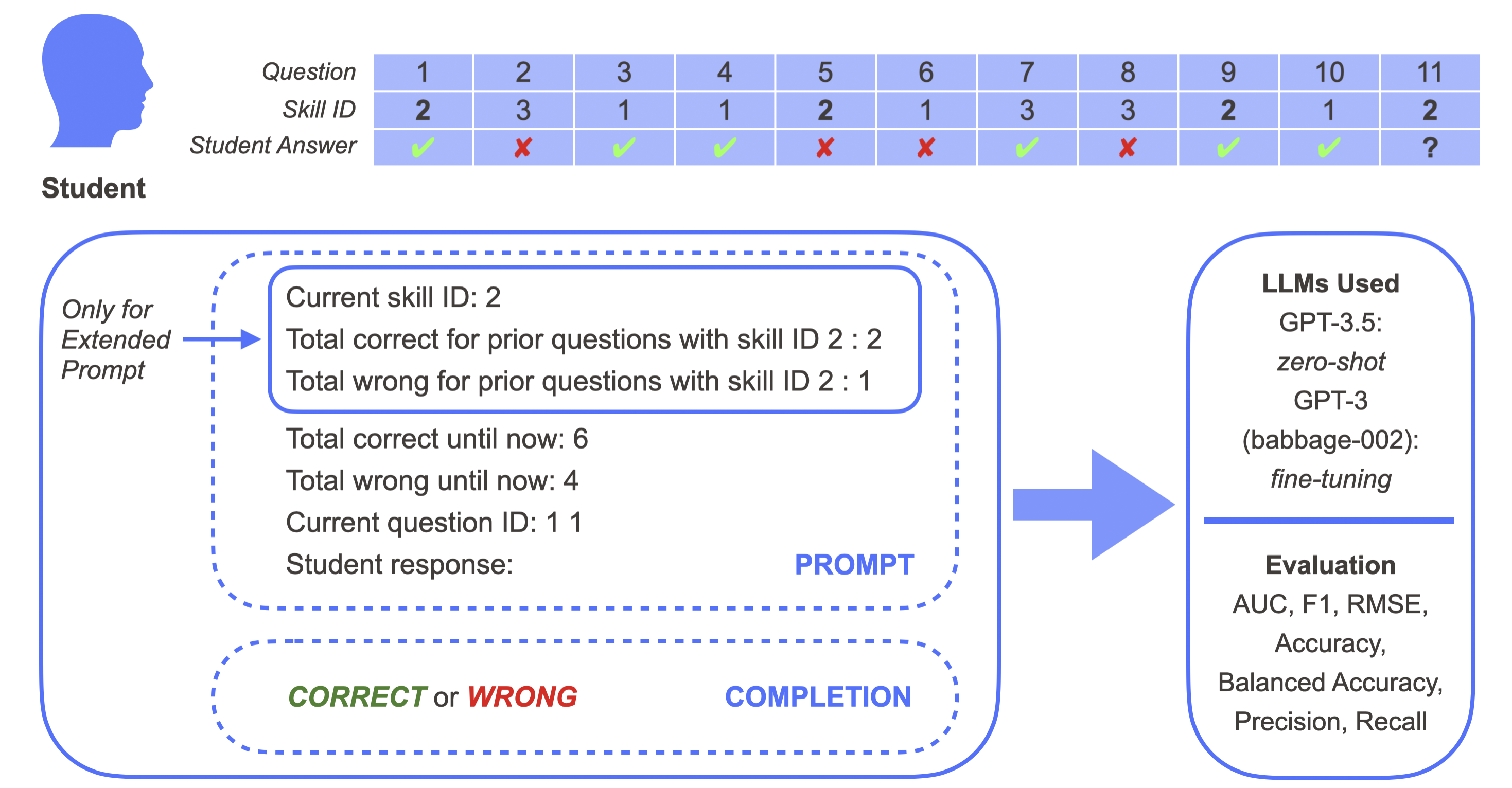}
    \caption{An overview of our prompting approach, following the feature sets used by \cite{gervet2020deep}, along with an example prompt based on a hypothetical sequence of student answers.}
    \label{fig:arch-diagram-kt}
    \Description{An overview of our prompting approach, along with an example prompt based on a hypothetical sequence of student answers.}
\end{figure*}

A diagram showing our prompting approach, along with an example, can be seen in Figure \ref{fig:arch-diagram-kt}.

\section{Methodology}
\label{sec:methodology}

\subsection{Datasets}

We evaluated the performance of the different models on three datasets with various sizes from the literature. For all datasets, we used the same train-test split as provided by \cite{gervet2020deep}.

\begin{itemize}
    \item \textbf{Statics:} This dataset, introduced by Steif and Bier \cite{steif2014oli}, is extracted from the data of an engineering statics course. It consists of 1223 unique problems and 282 unique users. Over all entries, 76.5\% (144883) are answered correctly and the rest (44414) are answered incorrectly.

    \item \textbf{ASSISTments 2009:} This dataset is compiled from the data collected using the ASSISTments system \cite{fengAddressingAssessmentChallenge2009}. It contains 17708 unique problems and 3114 unique users. Over all entries, 65.9\% (183303) are answered correctly and the rest (95033) are answered incorrectly.

    \item \textbf{ASSISTments 2017:} This dataset is also compiled from the data collected using the ASSISTments system \cite{fengAddressingAssessmentChallenge2009}. It contains 3162 unique problems and 1708 unique users. Over all entries, 37.4\% (349661) are answered correctly and the rest (584977) are answered incorrectly.
\end{itemize}

The Statics dataset is the smallest dataset (in terms of number of entries), and ASSISTments 2017 is the largest dataset, among the three datasets we used.

Before giving the datasets as inputs to our models, we discarded the entries of any student who had answered all questions correctly or all questions incorrectly. In this way, we only took into account the students who actively participated in a \textit{learning} process, rather than those who already knew the topic or did not learn it over time.
The removed students included 1.8\% of all students in Statics, 5.0\% of all students in ASSISTments 2009, and 0.0\% of all students in ASSISTments 2017.

\subsection{Evaluation and Metrics}
\label{ssec:eval-and-metrics}

We extracted the predicted responses of each of the models, along with their prediction probability, to compare the models we used in our downstream task. We used the following metrics, as defined and implemented in \texttt{scikit-learn} and previously used for evaluating models for the task of knowledge tracing \cite{sarsa2021empirical}, to evaluate our models (in all definitions, $TP$ = number of true positive predictions, $TN$ = number of true negative predictions, $FP$ = number of false positive predictions, and $FN$ = number of false negative predictions, where the CORRECT responses are considered \textit{positive} and the WRONG responses are considered \textit{negative}):

\begin{itemize}
    
    \item \textbf{Accuracy (Acc):} This metric is simply defined as the ratio of predictions matching with true labels, to the total count of predictions \cite{baratloo2015part}. It can be calculated using the expression $$\frac{TP+TN}{TP+TN + FP+FN}$$ as a number between zero and one.

    \item \textbf{Balanced Accuracy (Bal Acc):} We chose to report the \textit{balanced accuracy} metric as it takes into account the imbalance of the original data (regarding correct and wrong answers). This metric is defined as the arithmetic mean of \textit{specificity} and \textit{sensitivity}, and provided as a number between zero and one:

    $$\texttt{balanced-accuracy} = \frac{1}{2}\left( \frac{TP}{TP + FN} + \frac{TN}{TN + FP}\right )$$

    \item \textbf{Precision:} This metric is defined by the expression $$\frac{TP}{TP + FP}$$ as a number between zero and one.
    
    \item \textbf{Recall:} This metric is defined by the expression $$\frac{TP}{TP + FN}$$ as a number between zero and one.

    \item \textbf{F1:} This metric is defined as a \textit{harmonic} mean of precision and recall, and can be calculated using the expression $$\frac{2TP}{2TP + FP + FN}$$ as a number between zero and one.

    \item \textbf{Area Under Curve (AUC):} This metric is defined as the area under the Receiver Operating Characteristic (ROC) curve, as a number between zero and one.

    \item \textbf{Root Mean Square Error (RMSE):} This metric is defined as the root of the mean square error regression loss for prediction.

\end{itemize}

\section{Results}
\begin{figure}
    \centering
    \includegraphics[width=0.5\textwidth]{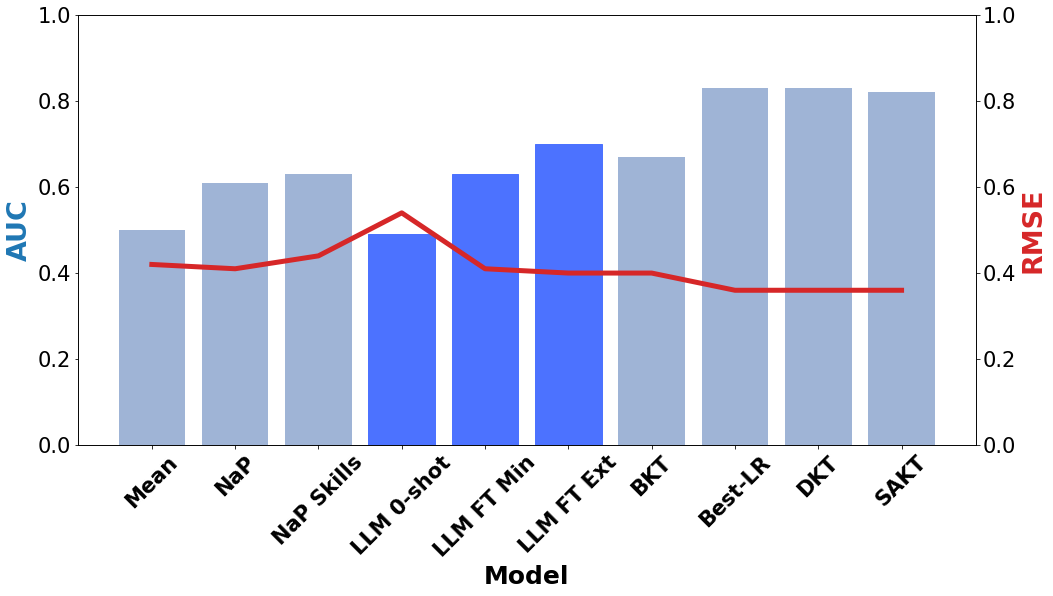}
    \caption{Comparing the AUC and RMSE scores of different models for Statics. LLM approaches are indicated in brighter blue.}
    \label{fig:plot-statics}
    \Description{Comparing the AUC and RMSE scores of different models for Statics.}
\end{figure}

\begin{figure}
    \centering
    \includegraphics[width=0.5\textwidth]{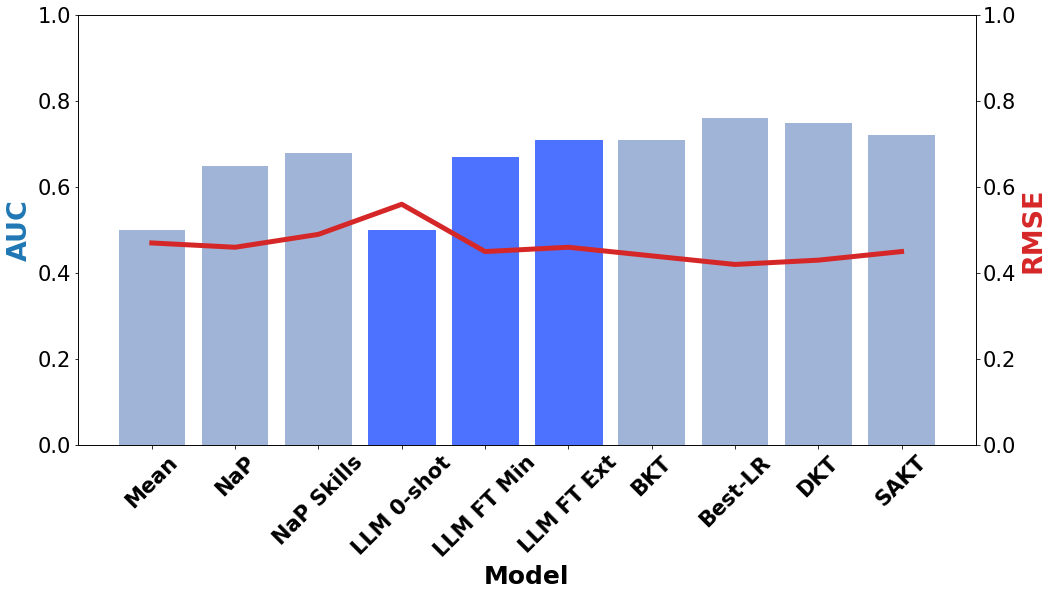}
    \caption{Comparing the AUC and RMSE scores of different models for ASSISTments 2009. LLM approaches are indicated in brighter blue.}
    \label{fig:plot-assistments2009}
    \Description{Comparing the AUC and RMSE scores of different models for ASSISTments 2009.}
\end{figure}

\begin{figure}
    \centering
    \includegraphics[width=0.5\textwidth]{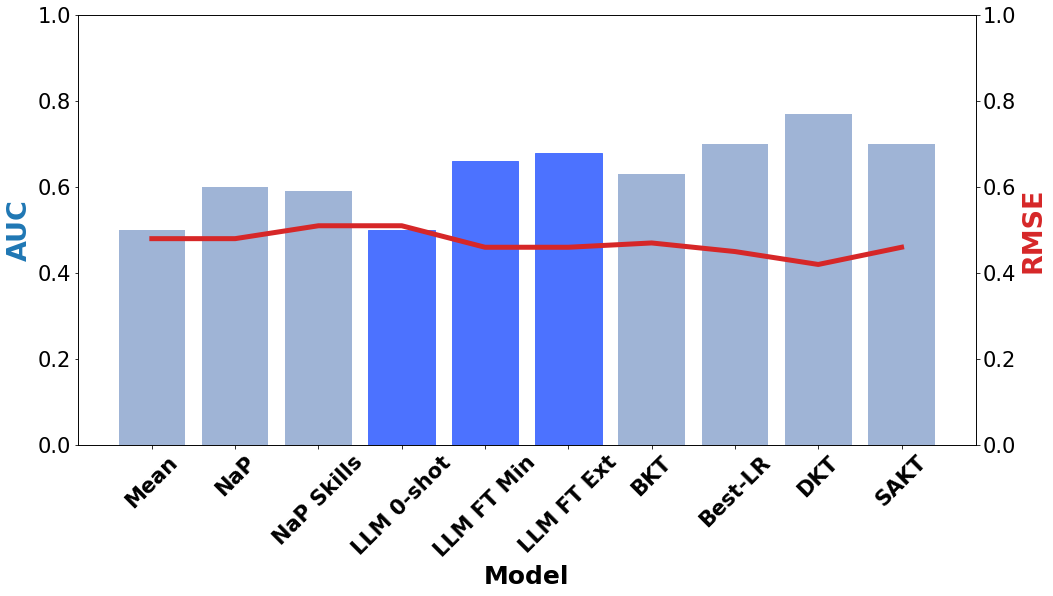}
    \caption{Comparing the AUC and RMSE scores of different models for ASSISTments 2017. LLM approaches are indicated in brighter blue.}
    \label{fig:plot-assistments2017}
    \Description{Comparing the AUC and RMSE scores of different models for ASSISTments 2017.}
\end{figure}

We provide an overview of our results on each of the datasets in each of the sections below.
First, we compare the two different LLM approaches, zero-shot prompting on GPT-3.5 and fine-tuning on GPT-3.
Then, we compare the LLM approaches to results from naive baselines.
Finally, we compare results between a set of widely used knowledge tracing models and the LLM approaches, with the goal of benchmarking the performance of the LLM approaches relative to the other approaches.
Full details containing results from all of our evaluation metrics across the three datasets can be found in Tables \ref{tab:statics}, \ref{tab:assist09}, and \ref{tab:assist17}. Plots comparing the AUC and RMSE scores across different models for each dataset can be seen in Figures \ref{fig:plot-statics}, \ref{fig:plot-assistments2009}, and \ref{fig:plot-assistments2017}.

\subsection{Comparing the Different LLM Methods}
When comparing the different LLM approaches, we found two general trends that persisted across all three datasets.
First, we found that fine-tuned LLMs dramatically outperformed the zero-shot method across metrics.
In particular, the best performing fine-tuned LLM on each dataset achieved AUC scores of 0.18 to 0.21 more than the zero-shot approach.
Then, when comparing the different fine-tuning approaches, we found a consistent but less obvious trend.
When the LLM was fine-tuned and evaluated on prompts that included the set of extended features, its performance was higher than the LLM which was fine-tuned on the minimal set of features.
The differences in AUC across the datasets ranged from 0.02 to 0.07, consistently in favor of the set of extended features.
The sole exception to this trend was on one dataset (ASSISTments 2009) where the minimal fine-tuning approach was competitive with the extended fine-tuning approach, in particular achieving higher balanced accuracy and lower RMSE.
Because of this second trend, when reporting results from the fine-tuning approaches in the rest of this section, we will only discuss the extended approach.

\subsection{Comparing the LLM Methods to Naive Baselines}

We observed that zero-shot approaches behaved worse than or equal to our naive baselines across all our datasets with regard to AUC.
On the other hand, we found that fine-tuning GPT-3 with the set of extended features as prompts beat all three naive baselines in the AUC score consistently across the three datasets.
The differences in AUC score between the fine-tuned models on extended prompts and our range of naive baselines ranged from 0.03 to 0.21, consistently in favor of the fine-tuned GPT-3 models.

\subsection{Comparing Fine-Tuned LLMs to Other Knowledge Tracing Models}
In comparing the GPT-3 models fine-tuned on the set of extended prompts to previously-used knowledge tracing approaches, we found three general trends, persistent across all of the three datasets.
First, we observed that the fine-tuned LLMs on the extended prompts achieved higher (for Statics and ASSISTments 2017) or similar (for ASSISTments 2009) AUC scores to the standard BKT model.
Second, we observed that other knowledge tracing models we used from previous works (DKT, Best-LR, and SAKT) consistently outperformed our fine-tuned LLMs in terms of AUC score.
The best performing models we tried from previous works were DKT and Best-LR, achieving AUC scores ranging from 0.04 to 0.13 above the fine-tuned LLMs on extended prompts across datasets.
Third, we found that the AUC scores of the fine-tuned LLMs on our extended prompts generally improved relative to the other knowledge tracing models as the size of the training data increased. 
For example, the AUC score of the fine-tuned LLM on extended prompts was only 0.02 below the Best-LR and SAKT models in ASSISTments 2017, while this difference changed to 0.13 and 0.12 when comparing LLMs fine-tuned on the Statics dataset to Best-LR and SAKT models, respectively.

\subsection{A Note on Fine-Tuning GPT-2 and BERT}
We opted not to present findings from fine-tuning GPT-2 or BERT, as neither model achieved the objectives of our task effectively. GPT-2, for instance, struggled to consistently predict "CORRECT" or "WRONG" and often produced other different tokens, hindering evaluation of its performance.

Regarding the text classification BERT model, although the fine-tuning process was successful in predicting outputs with the correct formatting due to the nature of the model, during training we noticed that the training loss plateaued and did not meaningfully decrease over successive iterations. Additionally, the predictions made on the test set were all constant values, consistently favoring the majority class over all predictions. As an example, for the Statics dataset, this naturally resulted in a recall score of 1.00 but a balanced accuracy of only 0.50. Given the constant values in its output, we deemed BERT ineffective for our task.

\begin{table*}
    \caption{Results for Statics dataset (Min = Minimal, Ext = Extended)}
    \centering
    \begin{tabular}{llccccccc}
    \hline
    \hline
         Family& Model&  AUC& F1& RMSE& Acc&  Bal Acc& Precision& Recall\\
         \toprule
         \multirow{3}{*}{Naive Baselines}
         &Mean& 0.50& \textbf{0.87}& 0.42& \textbf{0.77}& 0.50&  0.77&  1.00\\
         &NaP& 0.61& \textbf{0.87}& \textbf{0.41}& \textbf{0.77}& 0.50&  0.77&  1.00\\
         &NaP Skills& \textbf{0.63}& 0.83& 0.44& 0.73&  \textbf{0.55}&  0.79&  0.87\\
         \midrule
         \multirow{3}{*}{LLM}
         &0-Shot& 0.49& 0.48& 0.54& 0.41&  0.49&  0.76&  0.35\\
         % &0-Shot& 0.49& 0.49& 0.54& 0.47&  0.51&  0.79&  0.36\\
         % LLM FT Min&  0.66&  0.58&  0.59&  0.28&  0.38&  0.XX& 0.XX\\
         &FT Min& 0.63& \textbf{0.87}& 0.41& \textbf{0.77}&  0.52&  0.78&  0.98\\
         &FT Ext& \textbf{0.70}& \textbf{0.87}& \textbf{0.40}& \textbf{0.77}&  \textbf{0.53}&  0.78&  0.97\\
         \midrule
         Markov Model& BKT& 0.67& 0.87& 0.40& 0.78&  0.53&  0.78&  0.99\\
         Logistic Regression& Best-LR& \textbf{0.83}& \textbf{0.89}& \textbf{0.36}& \textbf{0.81}&  0.65&  0.83&  0.95\\         
         DL: RNN& DKT& \textbf{0.83}& 0.88& \textbf{0.36}& \textbf{0.81}&  \textbf{0.68}&  0.85&  0.93\\
         DL: Transformer& SAKT& 0.82& 0.88& \textbf{0.36}& \textbf{0.81}&  0.67&  0.85&  0.92\\
    \end{tabular}
    \label{tab:statics}
\end{table*}

\begin{table*}
    \caption{Results for ASSISTments 2009 dataset (Min = Minimal, Ext = Extended)}
    \centering
    \begin{tabular}{llccccccc}
    \hline
    \hline
         Family& Model& AUC& F1& RMSE& Acc&  Bal Acc& Precision& Recall\\
         \toprule
         \multirow{3}{*}{Naive Baselines}
         &Mean& 0.50& \textbf{0.80}& 0.47& 0.66&  0.50&  0.66&  1.00\\
         &NaP& 0.65&  0.79& \textbf{0.46}& \textbf{0.68}&  0.58&  0.70&  0.91\\
         &NaP Skills& \textbf{0.68}& 0.72& 0.49& 0.65&  \textbf{0.63}&  0.75&  0.70\\
         \midrule
         \multirow{3}{*}{LLM}
         &0-Shot& 0.50& 0.48& 0.56& 0.46&  0.50&  0.66&  0.37\\
         % &0-Shot& 0.50& 0.49& 0.56& 0.47&  0.51&  0.67&  0.38\\
         &FT Min& 0.67& 0.79& \textbf{0.45}& 0.68&  \textbf{0.60}&  0.71&  0.87\\
         &FT Ext& \textbf{0.71}& \textbf{0.81}& 0.46& \textbf{0.70}&  0.56&  0.69&  0.98\\
         \midrule
         Markov Model& BKT& 0.71& 0.80& 0.44& 0.71&  0.62&  0.73&  0.88\\
         Logistic Regression& Best-LR& \textbf{0.76}& \textbf{0.81}& \textbf{0.42}& \textbf{0.73}&  \textbf{0.66}&  0.75&  0.88\\
         DL: RNN& DKT& 0.75& \textbf{0.81}& 0.43& \textbf{0.73}&  \textbf{0.66}&  0.76&  0.87\\
         DL: Transformer& SAKT& 0.72& 0.78& 0.45& 0.70&  0.65&  0.75&  0.81\\
    \end{tabular}
    \label{tab:assist09}
\end{table*}

\begin{table*}
    \caption{Results for ASSISTments 2017 dataset (Min = Minimal, Ext = Extended)}
    \label{tab:assist17}
    \centering
    \begin{tabular}{llccccccc}
    \hline
    \hline
         Family& Model&  AUC& F1& RMSE& Acc&  Bal Acc& Precision& Recall\\
         \toprule
         \multirow{3}{*}{Naive Baselines}
         &Mean& 0.50& 0.00& \textbf{0.48}& 0.63&  0.50&  0.00&  0.00\\
         &NaP& \textbf{0.60}& 0.26& \textbf{0.48}& \textbf{0.65}&  0.55&  0.56&  0.17\\
         &NaP Skills& 0.59& \textbf{0.35}& 0.51& 0.63&  \textbf{0.56}&  0.50&  0.27\\
         \midrule
         \multirow{3}{*}{LLM}
         &0-Shot& 0.50& 0.30& 0.51& 0.57&  0.50&  0.37&  0.26\\
         &FT Min& 0.66& 0.38& \textbf{0.46}& 0.66&  0.58&  0.59&  0.28\\
         &FT Ext& \textbf{0.68}& \textbf{0.39}& \textbf{0.46}& \textbf{0.67}& \textbf{0.59}&  0.63&  0.28\\
         \midrule
         Markov Model& BKT& 0.63& 0.27& 0.47& 0.65& 0.55&  0.60&  0.17\\
         Logistic Regression& Best-LR& 0.70& 0.48& 0.45& 0.69&  0.63&  0.62&  0.39\\
         DL: RNN& DKT& \textbf{0.77}& \textbf{0.58}& \textbf{0.42}& \textbf{0.72}&  \textbf{0.68}&  0.64&  0.54\\
         DL: Transformer& SAKT& 0.70& 0.51& 0.46& 0.69&  0.63&  0.60&  0.43\\
    \end{tabular}
    \\
    \vspace{1em}
    The majority completion label in ASSISTments 2017, as opposed to the two other datasets, is "WRONG". As a result, the Mean baseline gives a constant output of 0, and thus, $TP = 0$, which in turn makes precision, recall, and F1 all to be zero (based on expressions in Section \ref{ssec:eval-and-metrics}).
\end{table*}

\section{Discussion}
The goal of this work was to explore whether LLMs are capable of modeling learner performance, and if they might offer a new approach to the task of predicting student behavior given prior activity.
In particular, we investigated the following research questions:
\begin{enumerate}
    \item Can the general pattern matching abilities of LLMs be applied to the domain of student performance prediction?
    \item How does the performance of LLM-based approaches on three real-world datasets compare to (a) naive baselines, (b) Markov approaches to knowledge tracing, (c) logistics regression approaches, and (d) deep learning approaches to knowledge tracing?
    \item Is there a relationship between the scale of a fine-tuned model (in terms of parameter count) and its ability to predict student performance?
\end{enumerate}

Regarding the first research question, our results indicated that LLMs are, in fact, capable of predicting student performance.
Even though the methods we developed were not able to achieve state-of-the-art results, our findings suggest that LLMs can serve as the basis for a new family of approaches to student performance modeling.
This potential is further underscored by our investigation into the second and third research questions, which revealed that the effectiveness of LLMs in this domain is influenced by a variety of factors.

Regarding the second research question, we found stark differences in performance between zero-shot and fine-tuning approaches across all three real-world datasets.
In particular, the zero-shot approach failed to beat any of the naive baselines on any meaningful metric, while the fine-tuned LLMs achieved higher AUC than all the naive baselines across all three datasets.
When compared against the other knowledge tracing models, the fine-tuned LLMs on the sets of extended features performed roughly on par with BKT across multiple metrics including AUC, balanced accuracy, F1, and RMSE.
Additionally, as the size of the dataset grew, the performance of the fine-tuned LLMs increased relative to the other knowledge tracing models that we evaluated.
On the two smaller datasets, Statics and ASSISTments 2009, the fine-tuned LLM performed roughly equal to BKT.
However, on the largest dataset, ASSISTments 2017, the fine-tuned LLM achieved higher AUC, lower RMSE, and higher balanced accuracy than BKT, and performed nearly as well as the other knowledge tracing models we used, with the exception of DKT.
These findings indicate that the accuracy of fine-tuned models improves as the volume of training data grows. However, a comprehensive analysis is essential to fully understand the dynamics of this relationship, suggesting a promising avenue for further research.

The third research question focused on the relationship between model scale and predictive performance in fine-tuned LLMs.
By comparing BERT and GPT-2 with GPT-3, we found evidence suggesting that the model scale has a notable impact on model performance in this domain.
% BERT and GPT-2 have similar numbers of parameters (109 million and 137 million respectively), roughly 1000 times smaller than GPT-3 (175 billion parameters).
The difference in scale and number of the model parameters of BERT and GPT-2 compared to GPT-3 corresponded to a dramatic difference in performance, with BERT's fine-tuned performance failing to beat the naive baselines, and with fine-tuned GPT-2 producing unusable output.
These findings suggest that modeling student performance may be an emergent ability \cite{weiEmergentAbilitiesLarge2022}, i.e., an ability that emerges in large language models but is not present in smaller models, and which cannot be straightforwardly inferred from the performance of the smaller models.
However, these findings say nothing about whether predictive performance will continue to improve with increasing parameter counts, highlighting a compelling topic for future work within the EDM community.

% From here: the new part added by Parsa
We found that fine-tuning GPT-3 consistently beat zero-shot prompting of GPT-3.5 across all three datasets, demonstrating that our zero-shot approach, which presented only one data point at a time for prediction with no memory built over time, struggled to identify patterns of student knowledge and learning.
A possible explanation for this difference has to do with the way that the knowledge tracing task is represented in the LLM regime.
In this regime, a single student's interaction data is represented as a sequence of natural language tokens, and the full dataset containing each student's data forms a corpus capturing a kind of specialized sub-language.
Modeling and prediction tasks thus rely on an LLM's ability to recognize or learn the underlying linguistic structure in the sub-language and extrapolate from seen to unseen data.
% LLMs are known to perform well on data they have observed in their training phase, even memorizing input data in certain cases \cite{de2023evaluation}.
Thus, one possible explanation for the difference in performance was that data points related to knowledge tracing and student modeling were underrepresented in the training dataset behind GPT-3.5.
This may also explain why fine-tuning GPT-3 was more effective, as it made it possible for the model to leverage its existing abilities to learn the linguistic structure of the knowledge tracing sub-language.
If this explanation is correct, our work shows that there exist natural language representations of the knowledge tracing task that allow an LLM to identify linguistic patterns of student behavior and to produce useful predictions about future performance.

\subsection{Limitations}
We have identified several limitations and areas of concern that warrant further discussion.
First, the absence of extensive publicly available details on the fine-tuning process used by OpenAI for GPT-3 presents a significant transparency issue.
Without access to the technical specifics of the exact fine-tuning method used by OpenAI, we are unable to fully understand the mechanisms through which LLMs can learn to predict student performance, or find if there are certain biases involved in the responses of the LLMs.
This issue underlines the importance of replicating these findings by fine-tuning an open-source LLM of comparable scale, such as LLaMA 2 \cite{touvronLlamaOpenFoundation2023}, to ensure a more transparent evaluation of the model's learning processes.

Second, because our methods do not explicitly model the dynamics of student learning or knowledge acquisition, their usefulness in intelligent tutoring systems may be limited.
One potential solution to this problem would be to use the fine-tuned model to make separate predictions about a student's future performance for each skill and to use these predictions to estimate a student's mastery level for each skill (similar to the method employed in \cite{gruverLargeLanguageModels2023}).
However, this approach would still fail to explicitly model the latent constructs of interest, and the numerous predictions needed at each timestep would be resource-intensive.

Relatedly, it is important to consider issues of sustainability when using LLMs in general, and especially in the context of student performance modeling.
While exact power estimates are unavailable, current estimates for a single LLM interaction are on the order of 1 Wh \cite{devriesGrowingEnergyFootprint2023}, which means that LLM-based approaches to student performance modeling use substantially more energy than other approaches.
While we do not feel this should hinder research on the use of LLMs for these types of tasks, we do feel that deployment of these models should be carefully considered until resource consumption can be brought to reasonable levels.

\subsection{Future Work}
In concluding our study, we identify several avenues for future research that build on our findings.
First, we found that incorporating more information into the prompts led to a modest improvement in the performance of fine-tuned models.
This observation calls for a more systematic investigation into the impact of including different features as prompts on model performance.
Possible features for exploration include the $k$-most recent responses across all questions, the $k$-most recent responses related to specific skills, the text of the actual questions, and the difficulty level of these questions.
This proposed work is crucial for identifying optimal strategies for prompt construction that could significantly enhance model effectiveness.

Second, despite the fact that our attempts employing a zero-shot approach did not yield successful outcomes, we hypothesize that a chain-of-thought prompting strategy \cite{weiChainofThoughtPromptingElicits2023} might offer a viable alternative.
Chain-of-thought prompting works by asking a model to first think step-by-step before answering, and has been shown to dramatically improve zero-shot performance on multiple tasks.
When combined with prompts containing rich sets of features about student performance and question characteristics, it is possible that an LLM would be able to accurately reason about a student's future performance.

Finally, our methodological approach involved condensing the history of student interactions into a concise prompt, drawing inspiration from logistic regression approaches.
This strategy effectively circumvented the limitations imposed by the LLM's context length constraints but came at the expense of creating in-context learning opportunities for the model.
This highlights another direction for future work, which is to investigate alternative strategies for incrementally feeding student interaction data into an LLM.
This would potentially allow the LLM to better engage with the temporal complexity of student interactions without breaching its context window limitations.

\section{Conclusion}
In this paper, we demonstrated that large language models (LLMs) are capable of modeling student performance and making useful predictions about future behavior.
Although our best-performing methods did not achieve state-of-the-art results on real-world datasets, they performed roughly on par with Bayesian Knowledge Tracing (BKT) and reliably beat a set of naive baselines.
These results suggest that LLMs may serve as the basis for a new family of approaches to knowledge tracing algorithms and student performance prediction.

\bibliographystyle{abbrv}
\bibliography{sigproc, knowledgetracing}

\balancecolumns
% That's all folks!
\end{document}